\documentstyle[preprint,aps]{revtex}
\begin{document}
\draft
\tightenlines
\title{Operator Formulation of Classical Mechanics and the Problem of
Measurement}
\author{Slobodan Prvanovi\'c \thanks{{\em E-Mail}: {\tt
prvanovic@phy.bg.ac.yu}} and Zvonko Mari\'c \thanks{{\em E-Mail}: {\tt
maric@shiva.phy.bg.ac.yu}}}
\address {Institute of Physics, P.O. Box 57, 11001 Belgrade, Serbia}
\maketitle
\begin{abstract}
 The basic concepts of classical mechanics are given in the
operator form. The dynamical equation for a hybrid system,
consisting of quantum and classical subsystems, is introduced and
analyzed in the case of an ideal nonselective measurement. The
nondeterministic  evolution is found to be the consequence of the
superposition of two different deterministic evolutions.
\end{abstract}
The correct theory of quantum mechanical (QM) and classical mechanical
(CM) systems in interaction has to differ from QM and CM in respect
to determinism and related topics. That is because the dynamical
equations of QM and CM can not lead to nonlinear changes of states
that can happen in the processes of (quantum) measurements. Quantum
and classical mechanics are deterministic theories in which pure states
can evolve only into pure states, not to the mixed ones. An approach
to a hybrid systems, a subsystem of which is QM system and another
one classical, given in \cite{1,2,3}, uses for states and observables
the direct product of QM and CM representatives. It was objected
\cite{2,3} that the dynamical equation used there does not save the
nonnegativity of states. This property of states has to be unaltered
if the theory is supposed to be physically meaningful (see page 331
of \cite{4}). We shall modify this approach by using the operator
formulation of CM. Then we shall introduce equation of motion for
hybrid systems and show that it appears to be capable of describing
the reduction (collapse) of states in the case of an ideal
nonselective measurement saving nonnegativity.

The most important features of classical mechanics, formulated in a
framework of real valued functions over phase space, are: {\bf 1.)}
the algebra of variables is the commutative one, {\bf 2.)} dynamical
equation is given by the Poisson bracket and {\bf 3.)} pure states
are those with sharp values of position and momentum the values of
which, in general, are independent. To show that this can be
formulated in the operator form, we shall proceed heuristicaly. Let
the pure states for a position, in Dirac notation, be $\vert q
\rangle $. Similarly, for a momentum $\vert p \rangle $. In quantum
mechanics independence of states is written by the use of the direct
product. With this prescription, pure classical states become $\vert
q \rangle \otimes \vert p \rangle$. So, let the space of states of
classical mechanics be the direct product of two rigged Hilbert
spaces ${\cal H}^q \otimes {\cal H}^p $ (to be precise, it would be a
subset of this). In such space, one can define commutative algebra of
classical observables as the algebra (over {\bf R}) of functions of
the operators $\hat q_{cm} \equiv \hat q \otimes \hat I$ and $\hat
p_{cm} \equiv \hat I \otimes \hat p$. States can be defined, like in
standard formulation, as functions of position and momentum, which
are now operators. That is, pure states are defined by:
$$
\delta (\hat q - q(t))\otimes \delta (\hat p - p(t)) = \int \int
\delta (q-q(t)) \delta (p-p(t)) \vert q \rangle \langle q \vert
\otimes \vert p \rangle \langle p \vert dqdp =
$$
\begin{equation}
=\vert q(t) \rangle \langle q(t) \vert \otimes \vert p(t) \rangle
\langle p(t) \vert ,
\end{equation}
while (noncoherent) mixtures are $\rho (\hat q_{cm} , \hat
p_{cm}, t)$.  These states are nonnegative and Hermitian operators
normalized to $\delta ^2 (0)$ if: $\rho (q,p,t) \in {\rm \bf R} $,
$\rho (q,p,t) \ge 0 $ and $\int \! \int \rho (q,p,t)\ dq\ dp=1$.
If one calculates the mean values by the Ansatz:
\begin{equation}
\langle f \rangle ={{\rm Tr} f(\hat
q_{cm},\hat p_{cm}) \rho (\hat q_{cm}, \hat p_{cm},t)\over {\rm Tr}
\rho (\hat q_{cm}, \hat p_{cm}, t) },
\end{equation}
then $\langle f \rangle $ will be equal to  standardly calculated
$\bar f = \int \! \int f(q,p) \rho (q,p,t)dqdp$.
The dynamical equation in operator formulation is defined as:
\begin{equation}
{\partial \rho (\hat q_{cm},\hat p_{cm},t) \over \partial t } 
={\partial H (\hat q_{cm},\hat p_{cm})
\over \partial \hat q_{cm}}
{\partial \rho (\hat q_{cm},\hat p_{cm},t) \over \partial \hat
p_{cm}} - {\partial H (\hat q_{cm},\hat p_{cm}) \over \partial \hat
p_{cm}} {\partial \rho (\hat q_{cm},\hat p_{cm},t) \over \partial
\hat q_{cm}}.
\end{equation}

The standard formulation of classical mechanics appears through the
kernels of the operator formulation in the basis $\vert q \rangle
\otimes \vert p \rangle$. This, together with (2), can be used as the
proof of equivalence of the two formulations. The prescription for
transition from c-number to operator formulation of CM consists in
simple change of real numbers $q$ and $p$ by Hermitian operators
$\hat q \otimes \hat I $ and $\hat I \otimes \hat p $.

To introduce appropriate formalism for hybrid systems, let us start
with standard treatment of two QM systems. When Hamiltonian is $\hat H
_{qm1} \otimes \hat H _{qm2}$, the states of these systems $\hat \rho
_{qm1}(t) \otimes \hat \rho _{qm2}(t)$ evolve according to
Schr\"odinger equation given by commutator for which it holds:
$$
{\partial (\hat \rho _{qm1}(t) \otimes \hat \rho _{qm2}(t)) \over
\partial t }=
{1\over i\hbar } [ \hat H _{qm1} \otimes \hat H _{qm2},
\hat \rho _{qm1}(t) \otimes \hat \rho _{qm2}(t) ] =
$$
\begin{equation}
= {1\over i\hbar } [ \hat H _{qm1} , \hat \rho _{qm1}(t) ] \otimes
\hat H _{qm2} \hat \rho _{qm2}(t) + \hat \rho _{qm1}(t) \hat H _{qm1}
\otimes {1\over i\hbar } [ \hat H _{qm2} , \hat \rho _{qm2}(t) ].
\end{equation}
Suppose now that the second system becomes classical. This would mean
that everything related to this system in (4) have to be translated
into classical counterparts. Having in mind the above formulation of
CM, we propose:
$$
{\partial (\hat \rho _{qm}(t) \otimes \hat \rho _{cm}(t) ) \over
\partial t }=
$$
\begin{equation}
= {1\over i\hbar } [ \hat H _{qm} , \hat \rho _{qm}(t)] \otimes
{\hat H _{cm} \hat \rho _{cm}(t) +\hat \rho _{cm}(t) \hat H
_{cm}\over 2} + {\hat H _{qm}  \hat \rho _{qm}(t)  + \hat \rho
_{qm}(t) \hat H _{qm} \over 2 }
\otimes \{ \hat H _{cm} , \hat \rho _{cm}(t)\},
\end{equation}
as dynamical equation. One can formally express (5) by ${\partial
\over \partial t} = {1\over
i\hbar } [\ \ \ \ , \ \ \ ] \otimes (\ \ \ , \ \ \ ) + (\ \ \ , \ \ \
) \otimes \{ \ \ \ , \ \ \ \}$, where $(\ \ \ , \ \ \ )$ stands for
symmetrized product. The explanation is as follows. The first system
remained QM so its type of evolution is left unaltered. Poisson
bracket is there instead of the second commutator because CM systems
evolve according to Liouville equation. It is defined as in (3), now
with partial derivatives in respect to $\hat q _{cm} \equiv \hat I
\otimes \hat q \otimes \hat I$ and $\hat p_{cm} \equiv \hat I
\otimes \hat I \otimes \hat p$. Both QM and CM states and observables
appear in the operator form, {\sl i.e.} hybrid system is defined in
${\cal H}_{qm} \otimes {\cal H}^q _{cm}  \otimes {\cal H}^p _{cm}$.
Without symetrization operators on the RHS of (5), in general, would
not be Hermitian for QM noncommutativity. Similar equations, in
the c-number formulation of CM, one can find in \cite{1,2,3}, where
it is antisymmetric, and in \cite{5}, where it is not.

The process of nonselective measurement can be considered within the
formalism of hybrid systems. In the case of an ideal measurement, the
state of the measured QM system and measuring apparatus (CM system)
evolves under the action of $ \hat H_{qm} \otimes \hat
I_{cm} + \hat I_{qm} \otimes \hat H_{cm} + \hat V _{qm} \otimes \hat
V _{cm}$, where, for instance, $\hat V _{qm} \equiv V_{qm} (\hat q
_{qm} , \hat p _{qm} )= V_{qm} (\hat q  \otimes \hat I \otimes \hat I
, \hat p  \otimes \hat I \otimes \hat I )$ and  $\hat V _{cm} \equiv
V_{cm} (\hat q _{cm} , \hat p _{cm} )= V_{cm} ( \hat I  \otimes \hat
q \otimes \hat I , \hat I  \otimes \hat I \otimes \hat p )$. The
measured observable is $\hat V_{qm} = \sum _i v_i \vert \psi _i
\rangle \langle \psi _i \vert \otimes \hat I \otimes \hat I $ and it
is necessary that $[\hat H_{qm} , \hat V_{qm} ]=0$ for if the quantum
system before the measurement was in one of the eigenstates of the
measured observable, say $\vert \psi _i \rangle$, does not
change the state during the measurement. Then $\hat H_{qm}$ can be
diagonalized in the same basis: $\hat H_{qm} = \sum _i h_i \vert \psi
_i \rangle \langle \psi _i \vert \otimes \hat I \otimes \hat I $. To
discuss the problem of measurement, there is the need to take for
the initial state of QM system the superposition of eigenstates of
the measured observable. The apparatus initially is in the state with
sharp values of position and momentum, {\sl i.e.}, the state of the
composite system at $t_o$ is $\hat \rho _{qm} (t_o) \otimes \hat \rho
_{cm} (t_o) = \sum _{ij} c_i (t_o) c_j ^* (t_o) \vert \psi _i \rangle
\langle \psi _j \vert \otimes \vert q_o \rangle \langle q_o \vert
\otimes \vert p_o \rangle \langle p_o \vert $. Substituting $\hat
\rho _{qm} (t) \otimes \hat \rho _{cm} (t)$ and each part of
Hamiltonian in ${\partial \over \partial t} = {1\over i\hbar } [\ \ \
\ , \ \ \ ] \otimes (\ \ \ , \ \ \ ) + (\ \ \ , \ \ \ ) \otimes \{ \
\ \ , \ \ \ \}$, one arrives to the expression for dynamics of
measurement:
$$
{\partial (\sum _{ij} c_{ij} (t) \vert \psi _i \rangle  \langle  \psi
_j \vert  \otimes \hat \rho ^{ij} _{cm}(t)) \over \partial t }=
\sum _{ij} c_{ij} (t) {1\over i\hbar } (h_i - h_j ) \vert \psi _i
\rangle  \langle  \psi _j \vert \otimes \hat \rho _{cm}^{ij} (t) +
$$
$$
+\sum _{ij} c_{ij} (t) \vert \psi _i
\rangle  \langle  \psi _j \vert \otimes \{ \hat H _{cm}, \hat \rho
_{cm}^{ij} (t) \} +
\sum _{ij} c_{ij} (t) {1\over i\hbar } (v_i - v_j ) \vert \psi _i
\rangle  \langle  \psi _j \vert \otimes
{1\over2} (\hat V_{cm} \hat \rho _{cm}^{ij} (t) +  \hat \rho
_{cm}^{ij} (t) \hat V_{cm} )+
$$
\begin{equation}
+\sum _{ij} c_{ij} (t) {1\over2} (v_i +v_j ) \vert \psi _i
\rangle  \langle  \psi _j \vert \otimes \{ \hat V _{cm}, \hat \rho
_{cm}^{ij} (t) \} .
\end{equation}
The last term on the RHS of (6) shows that the evolution of apparatus
depends on the eigenvalues of $\hat V _{qm}$ and, on the other hand,
as well as other terms, does not fix particular form of this
dependence. This is symbolically given by $\hat \rho _{cm}^{ij} (t)$.
For $i=j$ it is ease to solve (6) for CM parts of states:
$\hat \rho _{cm}^{ii} (t) = \vert q_i (t) \rangle \langle q_i (t)
\vert \otimes \vert p_i(t) \rangle \langle p_i(t) \vert $, where, for
instance, $\vert q_i (t) \rangle $ stands for $\vert q(v_i ,q_o ,p_o
,t) \rangle$. When the nondiagonal, $i\ne j$, terms are considered,
one can assume that CM parts of states depend on the eigenvalues of
$\hat V _{qm}$ in the same way as when $i=j$, {\sl i.e.}, one can
assume that a complete solution is:
\begin{equation}
\sum _{ij} c_{ij} (t) \vert \psi _i \rangle \langle \psi _j \vert
\otimes \vert
q_i (t) \rangle \langle q_j (t) \vert \otimes \vert p_i(t) \rangle
\langle p_j(t) \vert .
\end{equation}
But then partial derivatives ${\partial \over \partial \hat q_{cm}
}$ and ${\partial \over \partial \hat p_{cm} }$ would annihilate the
nondiagonal elements because they are not functions of $\hat q_{cm}$
and $\hat p_{cm}$ (they do not commute with $\hat q_{cm}$ and $\hat
p_{cm}$). There would be, for instance:
\begin{equation}
{\partial \over \partial \hat q } \vert q_i (t) \rangle \langle q_j
(t) \vert = {\partial \over \partial \hat q } \delta (\hat q - q_i
(t)) \delta _{i,j},
\end{equation}
and similarly for momentum. Consequently, for CM parts of the
nondiagonal elements the RHS of (6) would vanish, while on the LHS
there would not be zero. This contradiction implies that assumption
is not correct: solution of (6) can not be (7). Then, one can either
conclude that CM parts of nondiagonal terms do not depend on time or
one can assume that their time dependence differs from the assumed one.
In the first case, if one takes $\hat \rho _{cm}^{ij} (t)$ to be equal
to initial state $\vert q_o \rangle \langle q_o \vert \otimes \vert
p_o \rangle \langle p_o \vert $ for all $t$, then time derivative
would vanish, while the RHS of (6) would not be identically equal to
zero. This is the contradiction, too. Therefore, a solution of (6)
will be:
\begin{equation}
\sum _i \vert c_i (t_o) \vert ^2
\vert \psi _i \rangle \langle \psi _i \vert \otimes \vert q_i (t)
\rangle \langle q_i (t) \vert \otimes \vert p_i (t) \rangle \langle
p_i (t) \vert .
\end{equation}
In the second case, one should assume a solution, and find it, in the
form:
\begin{equation}
\sum _{ij} c_{ij} (t) \vert \psi _i \rangle
\langle \psi _j \vert \otimes \vert q_{ij} (t) \rangle \langle q_{ij}
(t) \vert \otimes \vert p_{ij} (t) \rangle \langle p_{ij} (t) \vert ,
\end{equation}
where, for instance, $\vert q_{ij} (t) \rangle \equiv \vert q({1\over
2} (v_i +v_j), q_o , p_o ,t) \rangle $.

It could be confirmed that both (9) and (10) are mixed states ($\hat
\rho ^2 \ne \hat \rho $) and that (7) is pure ($\hat \rho ^2 = \hat
\rho $). The state (7) can not be a solution of (6), due to (8),
while (9) and (10) do satisfy (6). Thus, it can be said that in
either case (6) produces nondeterministic (one to many) evolution
from initially pure state in (9) or (10). This is the crucial
difference between equation of motion for hybrid systems and
Schr\"odinger and Liouville dynamics that constitute it. Then, among
the last two variants, one has to decide by some further analyze
what would be the state of the measured system and apparatus.

The last operator, (10), is continuously related to the initial state
in respect to the form, while (9) is not. But, in difference to (9),
the solution (10) is meaningless for it is not nonnegative operator
(there could be the events with negative probability). The correct
way then to express the unavoidable transition from pure to mixed
state is by changing the form of a state, (9), not by changing
nonnegativity. Moreover, parts of nondiagonal terms in (10) are
regular states of CM system and they are accompanied with QM `states'
with vanishing trace. Such `states' only can be interpreted as
nonexisting. Regarding this, (10) becomes equal to (9) in which
nonexistence of this coherent terms is formulated in the proper
manner (the diagonal part is same anyhow). The dilemma of (9) and
(10) may be viewed in another way. When small QM system and big CM
system interact, there is the question which one would influence the
other. Either QM system would be forced to decohere, or CM one would
be driven in the same way by nonexisting something as by real QM
states. The second occasion has to be abandoned for it is,
metaphorically speaking, less possible than impossible (negative
probabilities are less than probability equal to zero which reflects
some impossibility). Perhaps it would be better then to say that (9)
is physical result of (6) and that (10) is physically unacceptable
mathematical solution.

The state (9) is in agreement with what is usually expected to happen
when the problem of measurement is considered in an abstract and
ideal form: To each eigenstate of the measured observable corresponds
one pointer position (and momentum). This occurs with probability
$\vert c_i (t_o) \vert ^2$ and takes place immediately after
apparatus in $\vert q_o \rangle \otimes \vert p_o \rangle $ has
started to measure $\hat V _{qm}$ on the system in $\sum _i c_i
(t_o) \vert \psi _i \rangle $.

If one takes (7) equal to the initial state for $t \rightarrow t_o$
(they both share the same characteristics) and compares it with (9),
then one can say that collapse is the consequence of partial
derivations which appear in Poisson bracket. Necessity of Poisson
bracket does not follow from the need to produce this discontinuous
evolution. On the contrary, it is necessary to give one to one
evolutions of CM subsystem when interaction term in Hamiltonian is
absent. There is nothing else in the equation of motion which can
be taken to be responsible for collapse; there are no {\sl ad hoc}
introduced projectors or stochastic interactions.

Once noticed departure from determinism in the formalism, it would be
noticed in (all) other aspects as some strange feature. For example,
in \cite{6} it was found that universal privileged times in
dynamics of hybrid systems appear. Here $t_o$ is such. In contrast to
opinion expressed there, we believe that this is rather nice property
of the approach. Namely, for described process, and all other that
can be treated in the same way, pure state can evolve to noncoherent
mixture, while noncoherent mixture can not evolve into coherent
mixtures - pure states, {\sl i.e.}, such processes are irreversible.
This means that for them the entropy only can rise or stay constant.
Then the distinguished moments of the rise of entropy can be used for
defining an arrow of time. It is interesting that nondeterministic
evolution of only CM system occurred in a treatment of CM by inverse
Weyl transform of the Wigner function \cite{7}.

We have discussed the form of solution of dynamical equation for
hybrid systems in the case of an ideal nonselective measurement.
Without an operator formulation of classical mechanics, the
argumentation would not be complete. It allowed us to consider
a solution as the pure state and to show that such solution can
not exist for the initial state of measured system being coherent
mixture of eigenstates of measured observable. Transition of the
apparatus from well defined initial state in appropriate pointer
positions has been analyzed in two versions. The common change of
purity in physically relevant case is followed, due to established
correlation between the apparatus and the system under observation,
by decoherence of quantum mechanical state. The nondeterministic
character of evolution comes from the superposition of two linear
dynamical equations. The reason for this lies in the fact that,
contrary to Schr\"odinger equation which is linear in respect to
both: the probabilities and the probability amplitudes, the operator
form of Liouville equation is linear only in respect to the
probabilities.


\begin{references}
\bibitem{1} I. V. Aleksandrov, Z. Naturforsch. {\bf 36a} (1981) 902
\bibitem{2} W. Boucher and J. Trashen, Phys. Rev. D Vol.{\bf 37}
No.{\bf 4} (1988) 3522
\bibitem{3} O. V. Prezhdo and V. V. Kisil, Phys. Rev. A Vol.{\bf 56}
No.{\bf 1} (1997) 162
\bibitem{4} A. Peres: Quantum Theory: Concepts and Methods, (Kluwer, 1993)
\bibitem{5} A. Anderson, Phys. Rev Lett. Vol.{\bf 74} No.{\bf 5}
(1995) 621
\bibitem{6} L. L. Salcedo, Phys. Rev. A Vol. {\bf 54} No. {\bf 4}
(1996) 3657
\bibitem{7} J. G. Muga and R. F. Snider, Europhys. Lett., Vol.{\bf
19} No.{\bf 7} (1992) 569
\end{references}
\end{document}